\documentclass[12pt]{article}
\usepackage{amsmath,amssymb,epsfig}
\makeatletter \@addtoreset{equation}{section} \makeatother
\renewcommand{\theequation}{\thesection.\arabic{equation}}
\newcommand{\ba}{\begin{array}}
\newcommand{\ea}{\end{array}}
\newcommand{\beq}{\begin{equation}}
\newcommand{\eeq}{\end{equation}}
\newcommand{\bea}{\begin{eqnarray}}
\newcommand{\eea}{\end{eqnarray}}

\def\bce{\begin{center}}
\def\ece{\end{center}}

\def\nonu{\nonumber}

\def\be{\beta}

\def\eps6{{\displaystyle \mathop{\epsilon}^{6}}{}}

\def\nab6{{\displaystyle \mathop{\nabla}^{6}}{}}

\def\0{{\sst{(0)}}}
\def\1{{\sst{(1)}}}
\def\2{{\sst{(2)}}}
\def\3{{\sst{(3)}}}
\def\4{{\sst{(4)}}}
\def\5{{\sst{(5)}}}
\def\6{{\sst{(6)}}}
\def\7{{\sst{(7)}}}
\def\8{{\sst{(8)}}}

\def\ba{\begin{array}}
\def\ea{\end{array}}
\def\beq{\begin{equation}}
\def\eeq{\end{equation}}
\def\be{\begin{equation}}
\def\ee{\end{equation}}

\def\eps{\epsilon}

\def\ba{\begin{array}}
\def\ea{\end{array}}
\def\beq{\begin{equation}}
\def\eeq{\end{equation}}
\def\be{\begin{equation}}
\def\ee{\end{equation}}

\def\eps{\epsilon}

\newcommand{\bean}{\begin{eqnarray*}}
\newcommand{\eean}{\end{eqnarray*}}

\begin{document}
\thispagestyle{empty} \addtocounter{page}{-1}
\begin{flushright}
KIAS-P07007 \\
{\tt hep-th/0702038}\\
\end{flushright}

\vspace*{1.3cm}

\centerline{ \Large \bf More on Meta-Stable Brane Configuration }
\vspace*{1.5cm}
\centerline{{\bf Changhyun Ahn} 
} 
\vspace*{1.0cm} 
\centerline{\it 
Department of Physics, Kyungpook National University, Taegu
702-701, Korea} 
\vspace*{0.8cm} 
\centerline{\tt ahn@knu.ac.kr} 
\vskip2cm

\centerline{\bf Abstract}
\vspace*{0.5cm}

We describe the intersecting brane configuration of type IIA string
theory corresponding to 
the meta-stable nonsupersymmetric vacua in four dimensional ${\cal
  N}=1$
supersymmetric $SU(N_c)$ gauge theory with   
an antisymmetric flavor, a conjugate symmetric flavor, 
eight fundamental flavors, $m_f$ fundamental flavors and $m_f$ 
antifundamental flavors.   
This is done by analyzing the ${\cal N}=1$ supersymmetric
$SU(2m_f-N_c+4)$
magnetic gauge theory with dual matters and 
the corresponding dual superpotential. 

\baselineskip=18pt
\newpage
\renewcommand{\theequation}
{\arabic{section}\mbox{.}\arabic{equation}}

\section{Introduction}

In the standard type IIA supersymmetric brane configuration  of
electric theory, the classical ${\cal N}=1$ SQCD with gauge group
$SU(N_c)$,
$N_f$ flavors of chiral superfields in the fundamental and
antifundamental
representations and vanishing superpotential can be described by
$N_c$ D4-branes suspended between NS5-brane and NS5'-brane along $x^6$
direction
together with $N_f$ D6-branes to the left of the NS5-brane, each of
which is connected to the NS5-brane by a D4-brane.
Similarly, in the standard type IIA 
supersymmetric brane configuration of magnetic theory,
 the classical ${\cal N}=1$ SQCD with gauge group
$SU(N_f-N_c)$,
$N_f$ flavors of chiral superfields in the fundamental and
antifundamental
representations and nonvanishing superpotential can be described by
$(N_f-N_c)$ color D4-branes suspended between NS5'-brane and NS5-brane along $x^6$
direction
together with $N_f$ D6-branes to the left of the NS5'-brane, each of
which is connected to the NS5'-brane by a flavor D4-brane \cite{EGK,GK}.

The dynamical supersymmetry breaking  \cite{ISS}
can occur in ${\cal N}=1$ SQCD with massive flavors where the
supersymmetry is broken by rank condition through F-term equations.
The geometric misalignment \cite{OO,FGU,BGHSS} 
of $N_f$ flavor D4-branes connecting between NS5'-brane
and D6-branes in the magnetic brane configuration can be interpreted
as nontrivial F-term equations in the corresponding gauge theory.
It turns out the minimal energy supersymmetry breaking IIA brane
configuration can be described by coincident
$(N_f-N_c)$ color 
D4-branes suspended between coincident $(N_f-N_c)$ D6-branes and NS5-brane along $x^6$
direction
together with the remaining coincident 
$N_c$ D6-branes to the left of the NS5'-brane, each of
which is connected to the NS5'-brane by a flavor D4-brane \cite{OO,FGU,BGHSS}.  
For symplectic or orthogonal gauge groups, similar analysis can be
done by adding orientifold 4-plane \cite{FGU,Ahn06-1}.

The classical ${\cal N}=1$ SQCD with gauge group
$SU(N_c)$,
$N_f$ flavors of chiral superfields in the fundamental and
antifundamental
representations, a single adjoint field 
and nonvanishing superpotential can be described by
$N_c$ color D4-branes suspended between two coincident
NS5-branes and NS5'-brane along $x^6$
direction
together with $N_f$ D6-branes  at values of $x^6$ that are between
those corresponding to the positions of the NS5-branes and NS5'-brane 
\cite{EGK,GK}.   
The dynamical supersymmetry breaking  \cite{AGM,AGM1}
can occur in ${\cal N}=1$ SQCD with massive flavors plus a single
adjoint by adding some mesonic deformation terms.
It turns out the minimal energy supersymmetry breaking IIA brane
configuration can be described by coincident
$(2N_f-N_c)$ color 
D4-branes suspended between coincident $(2N_f-N_c)$ D6-branes and
rotated two NS5-branes along $x^6$
direction,
the coincident 
$(N_c-N_f)$ D6-branes to the left of the NS5'-brane, each of
which is connected to the NS5'-brane by a flavor D4-brane,
together with the remaining coincident 
$(N_c-N_f)$ D6-branes to the left of the NS5'-brane, each of
which is connected to the NS5'-brane by a tilted flavor D4-brane \cite{Ahn06}.  

The classical ${\cal N}=1$ SQCD with gauge group
$SU(N_c)$,
$N_f$ flavors of chiral superfields in the fundamental and
antifundamental
representations, a symmetric flavor, a conjugate symmetric flavor 
and nonvanishing superpotential can be described by
$N_c$ color D4-branes suspended between  
rotated NS5-brane and rotated NS5'-brane along $x^6$
direction, an O6-plane,
together with $N_f$ D6-branes  at values of $x^6$ that are between
those corresponding to the positions of the rotated NS5-brane 
and NS5-brane(and their mirrors) \cite{LLL,GK}.   
It turns out the minimal energy supersymmetry breaking IIA brane
configuration can be described by coincident
$(2N_f-N_c)$ color 
D4-branes suspended between coincident $(2N_f-N_c)$ D6-branes and NS5-brane along $x^6$
direction
together with the remaining coincident 
$(N_c-N_f)$ D6-branes to the left of the NS5'-brane, each of
which is connected to the NS5'-brane by a flavor D4-brane(and their
mirrors) plus O6-plane \cite{Ahn07}.

In this paper, 
we study ${\cal N}=1$ $SU(N_c)$ gauge theory with 
an antisymmetric flavor $A$, a conjugate symmetric flavor $\widetilde{S}$, 
eight fundamental flavors $\hat{Q}$, $m_f$ fundamental
flavors $Q$ and $m_f$ antifundamental flavors $\widetilde{Q}$ 
\footnote{The proposal for the type IIA brane configuration describing the
nonsupersymmetric meta-stable minimum in this theory is given by 
\cite{FGU} already.} in the context of dynamical 
supersymmetric breaking vacua. 
The corresponding supersymmetric brane configuration in type IIA
string theory was given by various groups in \cite{LLL1,BHKL,EGKT} sometime ago.
See also \cite{LL}.
In the related gauge theory analysis \cite{ILS} alone, 
the turning on the superpotential 
$W =   (A \widetilde{S})^2$ was necessary to truncate the chiral
ring and the dual description was given. 
From the brane
analysis where, in general, two other quartic terms in the
superpotential are
present, 
this quartic superpotential can be obtained from
the corresponding ${\cal N}=2$ theory by integrating the massive
adjoint field out. 

For the limit of
infinite mass for the adjoint field,
the above quartic superpotential terms vanish because the coefficient
function appears as an inverse of mass. 
Now we can add the mass
term for quarks in the fundamental representation of the gauge group
while keeping the infinite mass limit. 
Then we turn to the
dual magnetic  gauge theories by standard brane motions with
appropriate linking number counting. It
turns out that the dual
magnetic theory giving rise to the meta-stable vacua 
is described by ${\cal N}=1$
$SU(2m_f-N_c+4)$ gauge theory with dual matter contents and the
superpotential.
Some of the F-term conditions cannot be satisfied by rank condition.
Then the final nonsupersymmetric minimal energy brane configuration for
this theory can be summarized in Figure 3
\footnote{There exists also other type of type IIA brane
  configuration \cite{ABFK} we do not discuss, 
describing the meta-stable vacuum in the 
${\bf Z}_2$ orbifold of the conifold. See also other brane
configuration in \cite{KOO}.}.  

In section 2, we review the type IIA brane configuration corresponding
to the electric theory based on the ${\cal N}=1$ $SU(N_c)$ gauge theory
with above various matter contents. 
In section 3, we construct the Seiberg dual magnetic theory which is 
${\cal N}=1$ $SU(3m_f-N_c+4)$ gauge theory with corresponding dual
matters as well as various gauge singlets, by brane motion and linking
number counting.
In section 4, we consider the infinite mass limit for the adjoint field of
${\cal N}=2$ theory in order to obtain nonsupersymmetric meta-stable
minimum where the gauge group is given by $SU(2m_f-N_c+4)$ with matter
contents
and present 
the corresponding intersecting brane configuration of type IIA string
theory.
In section 5, we describe M-theory lift of the supersymmetry breaking 
type IIA brane configuration we have considered in section 4.
Finally, in section 6, we give the summary of type IIA brane
configuration corresponding to meta-stable vacua of various gauge theories.

\section{The ${\cal N}=1$ supersymmetric electric brane configuration}

Let us describe ${\cal N}=1$ $SU(N_c)$ gauge theory with 
an antisymmetric flavor $A$, a conjugate symmetric flavor
$\widetilde{S}$, eight fundamental flavors $\hat{Q}$,  
$m_f$ fundamental flavors $Q$ and $m_f$ antifundamental flavors $\widetilde{Q}$
\cite{LLL1,BHKL,EGKT}. 
The anomaly-free global symmetry group determined by the gauge symmetry
on the D6-branes is given by 
$SU(m_f)_L \times SO(8)_L \times SU(m_f)_R $ with three $U(1)$'s
\footnote{Note that this gauge theory looks similar to \cite{ILS} in
  the sense that the matter contents are the same but the global
  symmetries and the superpotential are different. Over there, the
  global symmetry is $SU(m_f+8)_L \times SU(m_f)_R$ with three
  $U(1)$'s and the superpotential has more simple form. }. 
The type IIA 
brane configuration 
consists of three NS5-branes 
having different $x^6$ values, $N_c$
D4-branes 
suspended between them, $2m_f$ D6-branes, eight ``half'' D6-branes
extending all the way from $x^7=0$ to $x^7=\infty$, an
orientifold 6 plane($O6^{+}$-plane) located at negative $x^7$ carrying $+4$ RR charge and 
an
orientifold 6 plane($O6^{-}$-plane) located at positive $x^7$ carrying $-4$ RR charge
\footnote{The orientifold action identifies the two
factors of product gauge group $SU(N_c) \times SU(N_c)$ and projects the
bifundamental representation onto one antisymmetric tensor and one conjugate
symmetric tensor since there exists different projection acting on 
the different sides of the D4-branes \cite{BHKL}. Moreover, the matter
coming from the half D6-branes exists. 
The brane configuration of ${\cal N}=1$ supersymmetric
gauge theory with $SU(N_L) \times SU(N_R)$ and matter in the
bifundamental and fundamental representations was found in
\cite{BH,GP,ENR}.}. 

Let us introduce two complex coordinates, as usual,
\bea
v \equiv x^4 + i x^5, \qquad w \equiv x^8 + i x^9.
\nonu
\eea
According to ${\bf Z}_2$ symmetry of O6-plane (0123789) 
sitting at $v=0$ and $x^6=0$,
the coordinates $(x^4,x^5, x^6)$ transform as $-(x^4, x^5, x^6)$. 
Among three NS5-branes, the middle NS5'-brane (012389) which is located
at $x^7=0$ is stuck on an 
O6-plane and the two outer  NS5-branes are placed in a
${\bf Z}_2$ symmetric manner from ${\cal N}=2$ brane configuration as
in Figure 1. 
The left NS5-brane is rotated by an angle $\theta$ in
$(v,w)$ plane(denoted by $NS5_{\theta}$-brane) while 
the right NS5-brane is rotated by an angle $-\theta$
in the same plane(denoted by $NS5_{-\theta}$-brane).  
Our convention for this angle is different from the one used in \cite{LLL1}.
Similarly, the left $m_f$ D6-branes are rotated by an angle $\omega$ in
$(v,w)$ plane independently 
and denote them as $D6_{\omega}$-branes and the right $m_f$ D6-branes appear
as $D6_{-\omega}$-branes in a ${\bf Z}_2$ symmetric manner 
\footnote{There exists the figure showing the
  various angles in the
  product gauge group.  
One can make the following identifications explicitly : for the left branes 
$\theta=\frac{\pi}{2}-\theta_1$ and 
$\omega = \frac{\pi}{2}-\theta_1-\omega_1$ where $\theta_1$ and
$\omega_1$ are defined in the figure 5 of \cite{BHKL}.}. 
The middle NS5'-brane divides the O6-plane into two disconnected
regions corresponding to the positive and negative $x^7$,
 denoted by $O6^{-}$-plane and $O6^{+}$-plane respectively. The RR
charge of the O6-plane jumps from $+4$(in negative region of $x^7$) 
to $-4$(in positive region of $x^7$) as we cross the
NS5'-brane \cite{EJS}.
The part of $O6^{-}$-plane has more eight semi-infinite(or half)
D6-branes (0123789)
embedded in it.  
That is, four half D6-branes(and their mirrors) are present. 
This is necessary to preserve the charge conservation, or to cancel
the discontinuity of the RR charge \cite{HZ,LLL1,BHKL,EGKT}.  

For this brane setup, the classical superpotential in electric theory is
given by the quartic terms, flavor symmetry breaking term  and a mass
term for quarks 
as follows 
\footnote{If there exist $k$ coincident $NS5_{\theta}$-branes(and
  their mirrors also), then the superpotential
  takes the form  as $W= (\mu \Phi^{k+1} + A  \Phi
  \widetilde{S}+   Q \Phi \widetilde{Q}) +
  \hat{Q} \widetilde{S} \hat{Q} + m Q \widetilde{Q}$, in general.
  For the particular
  case where $NS5_{\theta}$-brane
  is parallel to the $D6_{\omega}$-branes(and their mirror also)
  indicating that there is no $ Q \Phi \widetilde{Q}$ term, 
one obtains $W = \frac{1}{2\mu} (A \widetilde{S})^{k+1} +  \hat{Q} \widetilde{S} \hat{Q}
  + m Q \widetilde{Q}$ \cite{BHKL}.  
\label{kgeneral}}
\bea
W = -\frac{1}{2 \mu} \left[ (A \widetilde{S})^2 +  Q \widetilde{S}
  A \widetilde{Q}
  + 
 (Q \widetilde{Q})^2\right] + 
 \hat{Q} \widetilde{S} \hat{Q}
+  m Q \widetilde{Q}, \qquad \mu \equiv \tan (\frac{\pi}{2}-\theta)
\label{electricsuperpotential}
\eea
where
$A$ is an antisymmetric field, $\widetilde{S}$ is a conjugate
symmetric field and $\hat{Q}$ is the 8 fundamental fields as well as 
$m_f$ fundamental fields $Q$ and $m_f$ antifundamental fields $\widetilde{Q}$
and the adjoint mass $\mu$ is related to the rotation angle $\theta$
as above \cite{Barbon,GK}.

Recall that starting with the brane configuration with ${\cal N}=2$
supersymmetry, one can rotate one or both of the NS5-branes or one or
more of the D6-branes such that the ${\cal N}=1$ supersymmetry is preserved.
The mass of the adjoint chiral superfield breaks the ${\cal N}=2$
supersymmetry to ${\cal N}=1$ supersymmetry. Keeping one of the
NS5-branes and all the D6-branes fixed and rotating the other
NS5-brane corresponds to changing the mass of the adjoint chiral
superfield.
On the other hand, rotating both NS5-branes relative to the D6-branes 
keeping the two NS5-branes and all the D6-branes parallel corresponds
to to changing the coupling between the adjoint chiral superfield and
quarks \cite{GK}.

When $NS5_{\theta}$-brane is parallel to $D6_{\omega}$-branes or $\theta=\omega$(and
their mirrors also), then the last two terms in the quartic terms
above (\ref{electricsuperpotential}) are not present because 
the coefficient function $\sin (\theta-\omega)$ appears in front of the $Q \Phi
\widetilde{Q}$ before integrating out the adjoint field $\Phi$
\footnote{When the $2m_f$
  D6-branes approach and are placed on top of the O6-plane, the tree level
  superpotential is given by $ (\mu \Phi^2 + A  \Phi
  \widetilde{S} + Q \Phi \widetilde{Q} )+( Q \widetilde{S} Q + 
\widetilde{Q} A \widetilde{Q})+ \hat{Q} \widetilde{S} \hat{Q}
  + m Q \widetilde{Q}$ where the global symmetry is enhanced
to $SO(2m_f+8)_L\times Sp(m_f)_R$. Note that the matter contents
in $Q$ and $\widetilde{Q}$ are doubled \cite{LLL1,BHKL,EGKT}. That is, 
an antisymmetric flavor $A$, a conjugate symmetric flavor
$\widetilde{S}$, eight fundamental flavors $\hat{Q}$,  
$2m_f$ fundamental flavors $Q$ and $2m_f$ antifundamental flavors $\widetilde{Q}$.
Instead, if we put $2m_f$
D6'-branes (0123457) approaching to the O6-plane, 
then the global symmetry becomes $SO(2m_f)_L \times
SO(8)_L \times Sp(m_f)_R$.   According to the result
  of \cite{EGKT,LLL}, the dual magnetic gauge group from brane
  configuration
is given by $SU(2m_f-N_c+4)$ and these extra two $\Phi$-independent terms in the
superpotential, $Q \widetilde{S} Q + 
\widetilde{Q} A \widetilde{Q}$, can be interpreted as mesonic perturbation in the
magnetic theory. \label{otherchiral} }.
We did not insert the dependence on this coefficient function into 
(\ref{electricsuperpotential}) explicitly in order to make it simple and actually
in the limit of $\mu \rightarrow \infty$(or $\theta \rightarrow 0$), 
all the quartic terms  vanish. 
The field theory analysis \cite{ILS} \footnote{When the baryonic
  operator $\widetilde{B}_n = \widetilde{S}^n \widetilde{Q}^{N_c-n}
  \widetilde{Q}^{N_c-n}$ 
gets an expectation value,
  then the initial gauge group $SU(N_c)$ along this flat direction 
is broken to $SO(n)$ with antisymmetric tensor $\hat{A}$ and $2m_f$
  vectors 
  as well as the singlets and the
  superpotential will be $ \hat{A}^2$ \cite{LLL1}. Along this flat
  direction,
the original theory flows to other theory in the IR \cite{BS}. Similarly, 
 when the baryonic
  operator $B_n = A^n Q^{N_c-2n}$ gets an expectation value,
  then the initial gauge group $SU(N_c)$ is broken to $Sp(n)$ with a conjugate
  symmetric tensor $\hat{\widetilde{S}}$ and $2m_f$ flavors  
and the
  superpotential will be $ \hat{\widetilde{S}}^2$.
\label{baryon}}
corresponding to the superpotential $(A \widetilde{S})^2$  which
was necessary to truncate the chiral ring  provides the magnetic dual description.
For the infinite limit of $\mu$, it is evident that the superpotential becomes 
$  \hat{Q} \widetilde{S} \hat{Q}+  m Q \widetilde{Q}$.
Since the eight half D6-branes are sitting on the top of O6-plane, the
flavor symmetry breaking term $ \hat{Q} \widetilde{S} \hat{Q}$ in the
superpotential indicates that the flavor symmetry $SU(m_f+8)_L$ is
broken to $SU(m_f)_L \times SO(8)_L$.

Then, the ${\cal N}=1$ supersymmetric electric brane configuration in
type IIA string theory realized in the gauge theory we study 
can be summarized as follows: 

$\bullet$
Middle NS5'-brane with worldvolume (012389) with $v=0=x^6=x^7$
\footnote{According to ${\bf Z}_2$ symmetry of O6-plane, there exist
  two possible orientations of a middle NS5-brane. It should be its
  own mirror under this action. It can extend in the $v$ direction
  given in \cite{Ahn07} or it can extend in $w$ direction. The second
  case provides the chiral theory \cite{LLL1}.  }.

$\bullet$
Left $NS5_{\theta}$-brane with worldvolume by both (0123) and two spatial dimensions
in $(v,w)$ plane and with negative $x^6$.

$\bullet$
Right $NS5_{-\theta}$-brane  with worldvolume by both (0123) and two spatial dimensions
in $(v,w)$ plane and with positive $x^6$.

$\bullet$ Left $m_f$
 $D6_{\omega}$-branes  with worldvolume by both (01237) and two spatial dimensions
in $(v,w)$ plane and with negative $x^6$. Before the rotation, the
 distance from D4-branes in the $v$ direction is nonzero.

$\bullet$ Right $m_f$
$D6_{-\omega}$-branes  with worldvolume by both (01237) and two spatial dimensions
in $(v,w)$ plane and with positive $x^6$.

$\bullet$ Eight semi-infinite
D6-branes  with worldvolume (0123789) with $x^7 \geq 0$ and $x^6=0$.
The four half D6-branes are mirror to the remaining four half
D6-branes with respect to O6-plane \footnote{Our convention for the
  location of these eight fundamentals are different from the one in
  \cite{GK} because the axis of $x^7$ is reversed.}. 

$\bullet$ Semi-infinite $O6^{+}$-plane  with worldvolume (0123789) with $v=0=x^6$
and $x^7 \leq 0$.

$\bullet$ Semi-infinite $O6^{-}$-plane  with worldvolume (0123789) with $v=0=x^6$
and $x^7 \geq 0$.

$\bullet$ $N_c$ D4-branes  with worldvolume (01236) with $v=0=w$.

We draw the type IIA electric brane configuration in Figure 1 which was basically given in
\cite{LLL1,BHKL,GK} already but the only difference is to put $D6_{\pm
\omega}$-branes in the
nonzero $v$ direction in order to obtain nonzero masses for the quarks
which are necessary to obtain the meta-stable vacua. 
Also one can imagine the brane configuration for 
$\theta=0$(in which the two outer $NS5_{\pm \theta}$ become two NS5-branes) from Figure 1.  

\begin{figure}[ht]
   \epsfxsize=4.5in 
\centerline{\epsffile{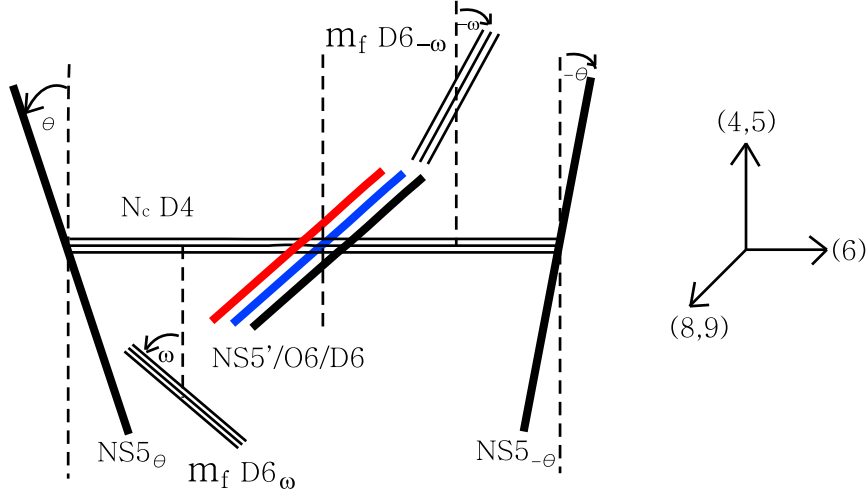}}
   \caption[FIG. \arabic{figure}.]{ 
The ${\cal N}=1$ 
supersymmetric electric brane configuration for the $SU(N_c)$ gauge
theory  with $A$,  
$\widetilde{S}$(that are strings stretching between D4-branes
located at the left hand side of O6-plane and those at the right hand
side of O6-plane), $\hat{Q}$(that are strings
stretching between D6-branes which are on top of $O6^{-}$-plane and D4-branes) and 
$Q, \widetilde{Q}$(that are strings
stretching between D6-branes and D4-branes).
The O6-plane is divided into half by the middle NS5'-brane in the
$x^7$-direction. 
The origin of
the coordinates $(x^6, v, w)$ is located at the intersection of
NS5'/O6/D6-branes(a combination of a middle
NS5'-brane, $O6^{+}$-plane, $O6^{-}$-plane and eight half D6-branes
above) 
and D4-branes. }
\label{fig1}
\end{figure}

\section{The ${\cal N}=1$ supersymmetric magnetic brane configuration}

In order to obtain the magnetic dual theory in the context of brane
configuration, 
we move the D6-branes and
NS5-branes through each other \cite{GK} and use the linking numbers \cite{HW} for the
computation of D4-branes which are created during this process.
In other words, the magnetic theory in the brane picture is obtained 
by interchanging D6-branes and NS5-branes with their mirror images 
while preserving the linking number.

The linking number \cite{HW} for the D6-branes is given by 
$
L_6 = \frac{1}{2}(n_{5L}-n_{5R}) +n_{4R}-n_{4L}
$
where $n_{5L,R}$ are the NS5-branes to the left or right of the
D6-branes and  $n_{4L,R}$ are the D4-branes to the left or right of the
D6-branes.
After we move the left $D6_{\omega}$-branes to the right all the way(and their
mirrors, right $D6_{-\omega}$-branes to the left) past the NS5'-brane and
$NS5_{-\theta}$-brane, 
the linking number $L_6$ of a $D6_{\omega}$-brane
becomes $L_6=1-n_{4L}$, 
which should be equal to the original 
$L_6=-1$  before the brane motion, 
according to the conservation of linking number. 
Then the $n_{4L}$ becomes 2 and we must add $2m_f$ D4-branes to the
left side of all $m_f$ $D6_{\omega}$-branes(and their mirrors). See
Figure 2 for the creation of these D4-branes. 

Next, we move the left $NS5_{\theta}$-brane to the right all the way past
O6-plane
(and its mirror, right $NS5_{-\theta}$-brane to the left), the linking number
of $NS5_{\theta}$-brane, 
$L_5 =\frac{m_f}{2}+\frac{1}{2} (4)-N+ 2 m_f$
using the formula $
L_5 = \frac{1}{2}(n_{6L}-n_{6R}) +n_{4R}-n_{4L}
$
where $n_{6L,R}$ are the D6-branes to the left or right of the
NS5-branes, $n_{4L,R}$ are the D4-branes to the left or right of the
NS5-branes. 
Originally, it was $L_5=-\frac{m_f}{2}-\frac{1}{2}(4)
+N_c$ from Figure 1 before the brane motion.
Therefore, we are left with the number of colors \cite{LLL1} in the magnetic theory
\bea
N = 3 m_f -N_c +4.
\nonu
\eea

The magnetic dual theory \footnote{In \cite{ILS}, the number of colors
is equal to $N=3m_f-12-N_c$ for $m_f$ fundamentals, $(m_f-8)$
antifundamentals, an antisymmetric tensor and a conjugate symmetric
tensor. By shifting as $m_f \rightarrow m_f+8$ in order to compare
with the gauge theory we consider and match with the matter contents, 
this becomes $N=3m_f+12-N_c$. The two
theories between the one in \cite{LLL1,BHKL} and the 
one in \cite{ILS} are connected to each
other. By giving the masses to all eight $\hat{Q}$'s in the
electric theory, the former theory will reduce to 
the latter theory.} 
has  gauge group $SU(N)$ where $N = 3 m_f
-N_c +4$ \footnote{When $k$ $NS5_{\theta}$-branes(and its mirrors) are
present, as in the notation of footnote \ref{kgeneral}, 
then by doing successive Seiberg dual, the magnetic dual
gauge group becomes $SU((2k+1)m_f -N_c +4(2k-1))$ where $2km_f$ term
comes from $k$ coincident $NS5_{\pm \theta}$-branes, $m_f$ term comes
from a middle NS5'-brane and $4(2k-1)$ constant term is due to the O6-plane.  }
and a conjugate symmetric $\widetilde{s}$ and an antisymmetric $a$
chiral multiplet together with $m_f$ fundamental $q$, 8 fundamental
$\hat{q}$ and $m_f$ antifundamental $\widetilde{q}$ chiral multiplets 
and moreover there exist gauge singlet fields
\bea
M_0 \equiv Q \widetilde{Q}, \qquad M_1 \equiv Q \widetilde{S} A
\widetilde{Q},
\qquad \widetilde{M} \equiv \hat{Q} \widetilde{Q} 
\label{mesons}
\eea
as well as
other type of gauge singlets $P \equiv Q\widetilde{S}Q $ 
and $\widetilde{P} \equiv 
\widetilde{Q} A \widetilde{Q}$ that are symmetric and antisymmetric in
their flavor indices respectively \footnote{
For the brane picture in the footnote  \ref{otherchiral},
after we move the left $NS5_{\theta}$-brane to the right all the way past
O6-plane(and its mirror, right $NS5_{-\theta}$-brane to the left), the linking number
of $NS5_{\theta}$-brane is given by 
$L_5 =\frac{(2m_f)}{2}+\frac{(4)}{2}-N$. Note that 
due to the doubling of the flavors of $Q$ and $\widetilde{Q}$, 
the number of D6-branes are $2m_f$
on top of O6-plane.
Originally, it was $L_5=-\frac{(2m_f)}{2}-\frac{(4)}{2}
+N_c$ before the brane motion. Then the magnetic gauge group one finds
is $SU(2m_f-N_c+4)$ as in footnote \ref{otherchiral}.
}.
Among newly created $2m_f$ D4-branes during the process from the electric theory to
magnetic theory, the $m_f$ D4-branes(when $m_f$ $D6_{\omega}$-branes 
cross a middle NS5'-brane)  in the magnetic theory can  split on the
$D6_{-\omega}$-branes in the most general way and a total of $m_f^2$
massless modes correspond to the $m_f^2$ components of the magnetic
meson $M_1$. On the other hand,  the remaining 
$m_f$ D4-branes(when $m_f$ $D6_{\omega}$-branes 
cross the right $NS5_{-\theta}$-brane)  in the magnetic theory 
can split similarly and $m_f^2$ massless modes correspond to 
the same components of the magnetic meson $M_0$. 
In the electric theory, if we move the $D6_{\omega}$-branes on top of
O6-plane, as in the footnote \ref{otherchiral}, we have seen the terms
like
as $P \equiv Q \widetilde{S} Q$ and $\widetilde{P} \equiv
\widetilde{Q} A \widetilde{Q}$. Here the
fundamental chiral multiplets $Q$ of the gauge group originate from 4-6
strings connecting D4-branes to $D6_{\pm \omega}$-branes ending a middle
NS5'-brane from positive $x^7$ while  the
antifundamental chiral multiplets $\widetilde{Q}$ of the gauge group come from 4-6
strings connecting D4-branes to $D6_{\pm \omega}$-branes ending a middle
NS5'-brane from negative $x^7$ \cite{EGKT,LLL1,BHKL,GK}.  
A 
meson $\widetilde{M}$ 
corresponds to 4-4 strings stretching 
between the right D4-branes and the left D4-branes.

The magnetic superpotential for arbitrary $\theta$ and $\omega$ 
is given by
\bea
W_{dual} = (a \widetilde{s})^2 +
M_1 q \widetilde{q} + M_0 q \widetilde{s} a \widetilde{q} +
P q \widetilde{s} q +
\widetilde{P} \widetilde{q} a \widetilde{q} + ( \hat{q} \widetilde{s}
\hat{q}+
\widetilde{M} \hat{q} \widetilde{q} ) + m M_0
\nonu
\eea
where the first five terms are obtained from the field theory analysis
\cite{ILS},
the next two terms are due to the presence  of flavor symmetry
breaking (\ref{electricsuperpotential}) in the electric theory and the
last term comes from the mass of quarks in the electric theory
\footnote{As for the brane configuration in footnote
  \ref{otherchiral}, by adding the composite operators $Q
  \widetilde{S} Q$ and $\widetilde{Q} A \widetilde{Q}$ in electric theory,
one gets the generalized magnetic superpotential by adding $P$- and
$\widetilde{P}$-terms into $W_{dual}$, observed in \cite{EGKT}.  }.
In particular, $P$ and $\widetilde{P}$-terms arise when
the $m_f$ $D6_{\omega}$-branes intersect with its mirror 
$m_f$ $D6_{-\omega}$-branes around the location of a middle NS5'-brane. 
In next section, by considering the subset of this process from the
electric theory to magnetic theory, we'll show how the brane
configuration
for the meta-stable supersymmetry breaking states arises.

We draw the type IIA magnetic brane configuration in Figure 2 which was given in
\cite{LLL1,BHKL,GK} already but the only difference is to put $D6_{\pm
\omega}$-branes in the
nonzero $v$ direction in order to obtain nonzero masses for the quarks
which are necessary to obtain the meta-stable vacua. 
Compared with the previous Figure 1, 
this Figure 2 looks similar to Figure 1 but 
there exist $2m_f$ D4-branes(and its mirrors) which will be reduced to
$m_f$ D4-branes(and its mirrors) for infinite adjoint mass limit in
next section. 
Also one can imagine the brane configuration for 
$\theta=0$(in which the two outer $NS5_{\pm \theta}$ become two
NS5-branes) 
from Figure 2.  

\begin{figure}[ht]
   \epsfxsize=4.5in 
\centerline{\epsffile{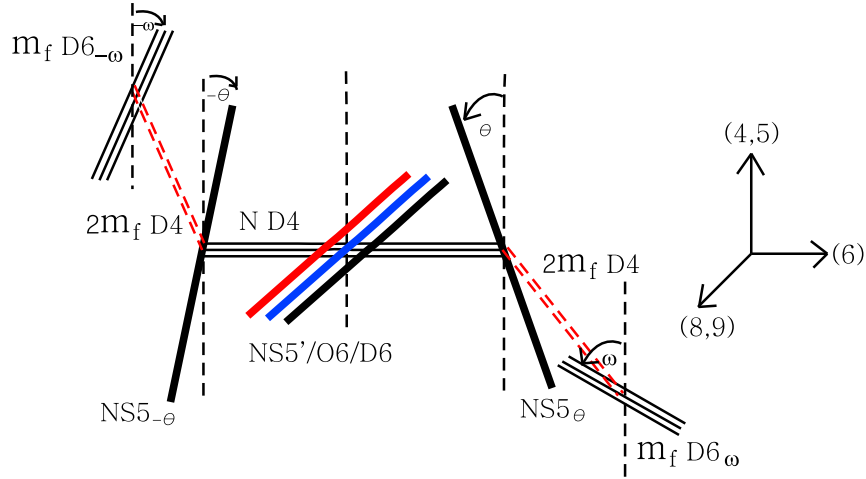}}
   \caption[FIG. \arabic{figure}.]{ 
The ${\cal N}=1$ supersymmetric magnetic 
brane configuration for the $SU(N=3m_f-N_c+4)$ gauge
theory  with $a$, 
$\widetilde{s}$, $\hat{q}$, 
$q$ and $\widetilde{q}$.
 The $2m_f$ D4-branes connecting between
$NS5_{-\theta}$-brane and $D6_{-\omega}$-branes are the dual gauge singlets
corresponding to the mesons $M_0$ and $M_1$ (\ref{mesons}). 
As will be explained in section
\ref{four},
in the $\mu \rightarrow \infty$ limit (or $\theta \rightarrow 0$ limit),
only $m_f$ D4-branes connecting NS5-brane and D6-branes appear(and
their mirrors). In the superpotential, it contains either $M_0$ or
$M_1$-dependent term depending on which way the brane motion occurs.}
\label{fig2}
\end{figure}

\section{Nonsupersymmetric meta-stable brane configuration
\label{four} }

When the D6-branes are parallel to a middle NS5'-brane or 
perpendicular to NS5-branes in the electric theory(in other words, the
angles are given by $\omega=\frac{\pi}{2}$ and $\theta=0$) in the $\mu
\rightarrow \infty$ limit
\footnote{In this case, the D4-branes can move freely in $v$ direction
along the two outer NS5-branes and $A$ and $\widetilde{S}$ are free to
get vacuum expectation values \cite{ILS,BHKL}.},
let us describe the nonsupersymmetric meta-stable brane configuration 
\footnote{For the case of $\theta=0=\omega$(when  the D6-branes are
  perpendicular to 
a middle NS5'-brane), the superpotential contains $M_1$-dependent term
because the $m_f$ D4-branes created during the Seiberg dual process
correspond to the singlet $M_1$, as in \cite{Ahn07}. }.
First rotate $D6_{\omega}$-branes to the $v$ direction and make them to be
parallel to a middle NS5'-brane(and its mirrors also).  
After we move the left $D6_{\omega=\frac{\pi}{2}}$-branes to the right all the way(and their
mirrors, right $D6_{-\omega=-\frac{\pi}{2}}$-branes 
to the left) past the middle NS5'-brane and
the right NS5-brane(=$NS5_{-\theta=0}$-brane), 
the linking number $L_6$ of a single D6-brane 
becomes $L_6=\frac{1}{2}-n_{4L}$, 
which should be equal to the original linking number 
$L_6=-\frac{1}{2}$. Let us denote
$D6_{\omega=\pm\frac{\pi}{2}}$-branes 
by unrotated D6-branes (0123789). 
Then we should add $m_f$ D4-branes, corresponding to the meson $M_0$, to the left 
side of all the right $m_f$ D6-branes(and their mirrors) in Figure 2. 
Note that the difference, compared with previous case where $0<\theta
\leq \frac{\pi}{2}$, 
appears here when a D6-brane crosses the middle NS5'-brane.
Due to the parallelness of these(D6-branes and NS5'-brane), 
there is no creation of D4-branes. This fact will affect the number of
color
D4-branes below. 

Next, we move the left NS5-brane(=$NS5_{\theta=0}$-brane) to the right all the way past
O6-plane(and its mirror, right NS5-brane to the left), and then the linking number
of NS5-brane becomes 
$L_5 =\frac{m_f}{2}+\frac{1}{2}(4) -N +  m_f$. Note that 
since the NS5-brane is nonparallel to the O6-plane(actually they are
perpendicular to each other), one can see the contribution from the
O6-plane which plays the role of four D6-branes, contrary to the
result of \cite{Ahn07} in which the outer NS5'-branes were parallel to
an O6-plane.
Originally, the linking number was $L_5=-\frac{m_f}{2}-\frac{1}{2}(4) +N_c$.
This implies that the number of D4-branes in magnetic theory, $N$, becomes 
\bea
N = 2 m_f -N_c + 4 
\nonu
\eea
which is the same as the case for the theory of the footnote  \ref{otherchiral}.

The magnetic superpotential corresponding to the electric 
superpotential 
$W=  \hat{Q} \widetilde{S} \hat{Q}+  m Q \widetilde{Q}$ with $\mu
\rightarrow \infty$
is given by
\bea
W_{dual} = M_0 q \widetilde{s} a \widetilde{q} + m M_0
 + ( \hat{q} \widetilde{s}
\hat{q}+
\widetilde{M} \hat{q} \widetilde{q} ). 
\label{finalsuper}
\eea
When we take $\mu \rightarrow \infty$, the quartic terms from 
(\ref{electricsuperpotential}) vanish and this leads to the fact that 
there is no such term like $(a\widetilde{s})^2$ in the magnetic
superpotential
we are considering.  
Our particular route for brane motion, as we observed before, 
does not allow for us to have $M_1$ term which also 
prohibits the presence of $M_1 q \widetilde{q}$.
Moreover since the D6-branes are parallel to the middle 
NS5'-brane($\omega=\frac{\pi}{2}$), there are no $P$ or
$\widetilde{P}$ dependent terms.
Eventually, we are left with the result of (\ref{finalsuper}).  

Here $q$ and $\widetilde{q}$ are fundamental and antifundamental for
the gauge group indices and antifundamentals for the flavor indices.
The fields $a$ and $\widetilde{s}$ are antisymmetric tensor and conjugate symmetric
tensor for the gauge group indices respectively with no flavor
indices.
Then, $q \widetilde{s} a \widetilde{q}$ has rank $N$ while $m$ has a
rank $m_f$.  Therefore, the F-term condition, the derivative the 
superpotential $W_{dual}$ with respect to $M_0$, cannot be satisfied 
if the rank $m_f$ exceeds $N$. This is so-called rank condition.    
Other F-term equations are satisfied by taking the vacuum expectation 
values of $\widetilde{M}$ and $\hat{q}$ to vanish.

The classical moduli space of vacua can be obtained from F-term
equations.
From the F-terms $F_{q}$ and $F_{\widetilde{s}}$, one gets
$\widetilde{s} a \widetilde{q} M_0 =0=   a \widetilde{q} M_0 q +
\hat{q} \hat{q}$.
Similarly, one obtains 
$\widetilde{q} M_0 q \widetilde{s} =0=   M_0 q \widetilde{s} a+
\widetilde{M} \hat{q}$ from 
the F-terms $F_{a}$ and $F_{\widetilde{q}}$. Moreover, there is a relation 
$ q \widetilde{s} a \widetilde{q} +  m=0$ from the F-term $F_{M_0}$.
From the conditions $a \widetilde{q} M_0=0= M_0 q \widetilde{s}$ and 
$\widetilde{M}=0=\hat{q}$, which
satisfy the above four equations for nonzero vacuum expectation
values $q, \widetilde{q}, a$ and $\widetilde{s}$, one
can fix the form of $M_0$ and part of $q \widetilde{s}$ and $a
\widetilde{q}$, which can be fixed by using $F_{M_0}=0$ further, and one obtains 
the following solutions 
\bea
<q \widetilde{s}> =  \left(
\begin{array}{c}
\sqrt{m} e^{\phi} {\bf 1}_{N}  \\
0
\end{array}
\right),  
<a \widetilde{q}> =
 \left(
\begin{array}{cc}
\sqrt{m} e^{-\phi}  {\bf 1}_{N}   &
0
\end{array}
\right), 
<M_0>  =
 \left(
\begin{array}{cc}
0  & 0 
 \\
0 & \Phi_0  {\bf 1}_{m_f-N} 
\end{array}
\right)
\label{vacuum}
\eea
as well as $<\widetilde{M}>=0=<\hat{q}>$
where $\Phi_0  {\bf 1}_{m_f-N}$ is an arbitrary 
$(m_f-N) \times (m_f-N)$ matrix and the zeros of 
$< q \widetilde{s}>$ and $< a \widetilde{q}> $ are $(m_f-N) \times N $
and $N \times (m_f-N)$ zero matrices respectively. Similarly, the zeros
of $m_f \times m_f$  matrix $M_0$ are assumed also.
Then $\Phi_0$ and $(\sqrt{m} e^{\phi}, \sqrt{m} e^{-\phi})$
parametrize
a pseudo-moduli space.
Let us expand around on a point on (\ref{vacuum}), as done in \cite{ISS}. 
That is, 
\bea
q  & = &
\left(
\begin{array}{c}
q_0  {\bf 1}_{N} +\frac{1}{\sqrt{2}}(\delta \chi_{+} + \delta \chi_{-})
 {\bf 1}_{N} \nonu \\
\delta \varphi
\end{array}
\right), 
\widetilde{s} = \left(\widetilde{s}_0 + \delta
  \widetilde{X} \right) {\bf 1}_{N}, 
a   =   \left(a_0 + \delta
  X\right) \sigma_2 \otimes {\bf 1}_{\frac{N}{2}},
\nonu \\
\widetilde{q}  & = &
 \left(
\begin{array}{cc}
\widetilde{q}_0    \sigma_2 \otimes {\bf 1}_{\frac{N}{2}} +
\frac{1}{\sqrt{2}}(\delta \chi_{+} - \delta \chi_{-})  \sigma_2 \otimes
  {\bf 1}_{\frac{N}{2}}   &
\delta \widetilde{\varphi}  \sigma_2 \otimes  {\bf 1}_{\frac{m_f-N}{2}} 
\end{array}
\right), 
M_0  =
 \left(
\begin{array}{cc}
\delta Y  & \delta Z
 \\
\delta \widetilde{Z} & \Phi_0  {\bf 1}_{m_f-N} 
\end{array}
\right)
\nonu
\eea
together with the variations of $\widetilde{M}$ and $\hat{q}$ that are
$m_f \times 8$ and $8 \times N$ matrices respectively.
Then the superpotential (\ref{finalsuper}) becomes 
\bea
W_{dual}^{fluct} & = &  \Phi_0 \left( \delta \varphi \; \widetilde{s}_0 
\; a_0 \; \delta \widetilde{\varphi}  + m \right) +
  \delta Z \; \delta \varphi \; \widetilde{s}_0 \; a_0 \; \widetilde{q}_0
+  \delta \widetilde{Z} \; q_0 \; \widetilde{s}_0 \; a_0 \; 
\delta \widetilde{\varphi}  
\nonu \\
 & + & \left( \frac{1}{\sqrt{2}}
\delta Y \; \delta \chi_{+} \; \widetilde{s}_0 \; a_0 \; \widetilde{q}_0  
+ \cdots \right)
+ \mbox{(cubic)}
\label{s}
\eea
where $\mbox{(cubic)}$ stands for the terms that are cubic or higher
in the fluctuations and $\cdots$ contains some parts from the last two
terms in (\ref{finalsuper}) that are not relvant.
Then to quadratic order, the model splits into two sectors where the
first piece given by the first line of (\ref{s}) is an O'Raifeartaigh type model 
and the  second piece given by the second line of (\ref{s})
is supersymmetric and will not contribute to the
supertrace. The fields 
$\delta \chi_{\pm}, \delta Y, \delta X$ and $\delta \widetilde{X}$
couple to the supersymmetry breaking fields $\delta \varphi$ and
$\delta \widetilde{\varphi}$ via terms of cubic and higher
 order in the fluctuations.
Then, the remaining relevant terms of superpotential are given by
\bea
W_{dual}^{rel} & = &  \Phi_0 \left( \delta \hat{\varphi}  
\; \delta \hat{\widetilde{\varphi}} + m \right) +
  \delta Z \; \delta \hat{\varphi} \; a_0 \; \widetilde{q}_0 
+ \delta \widetilde{Z} \; q_0 \; \widetilde{s}_0 \;
\delta \hat{\widetilde{\varphi}}
\nonu
\eea
where 
$ \delta \hat{\varphi} \equiv \delta \varphi \; \widetilde{s}_0 $ and 
$\delta \hat{\widetilde{\varphi}} \equiv a_0 \; \delta 
\widetilde{\varphi} $.
At one loop, the effective potential $V_{eff}^{(1)}$ for $\Phi_0$ 
can be obtained 
from this relevant part of superpotential which consists of 
the matrix $M$ and $N$ of \cite{Shih} 
where the defining function ${\cal F}(v^2)$ can be
computed.
Using the equation (2.14) of \cite{Shih} 
of $m_{\Phi_0}^2$ and ${\cal F}(v^2)$, one gets 
that $m_{\Phi_0}^2$ will contain $(\log 4 -1) > 0$, by
taking the limit where $q_0 \widetilde{s_0} \rightarrow \sqrt{m}
e^{\phi}$
and  $a_0 \widetilde{q}_0 \rightarrow \sqrt{m}
e^{-\phi}$, as in (\ref{vacuum}).
This implies that these vacua are stable.
In the brane configuration from Figure 2, since the $N$ D4-branes 
can slide along the two NS5-branes when $\theta=0$, 
the vacuum expectation values
of
$a$ and $\widetilde{s}$ can be $2 \times 2$ block-diagonalized. 

By recombination of $N$ of D4-branes connecting D6-branes and
NS5-brane with those connecting NS5'-brane and NS5-brane and moving
them in $v$ direction(and their mirrors) from Figure 2, 
the minimal energy supersymmetry breaking brane
configuration is shown in Figure 3 
that was observed also in \cite{FGU}.
This phenomenon occurs all the examples found in 
\cite{OO,FGU,BGHSS,Ahn06,Ahn06-1,Ahn07}.
Compared with the brane configuration of \cite{Ahn07},
the locations of NS5-brane and NS5'-brane are interchanged each other.
Roughly speaking, when the NS5-brane moves along the $x^6$ direction
and two NS5'-branes move to the origin from the Figure 3 of
\cite{Ahn07}, 
then one obtains the Figure 3 of this paper. Undoing the Seiberg dual
in the context of \cite{FGU}. 
Some entries in the dual quarks $q, \widetilde{q}$ and dual tensors
$a, \widetilde{s}$ acquire the nonzero expectation values in terms of
the eigenvalues of mass matrix $m$ to minimize
the F-term $F_{M_0}$ in this dual gauge theory. 
This corresponds exactly to the geometric recombination 
and movement along the $v$ direction above. 
When one gets the nonzero vacuum expectation values for $<M_0>, <q>$, and
$<\widetilde{q}>$, then 
the separation of the D4-branes along the middle NS5'-brane  
corresponds to the mass of two index tensor $(a \widetilde{s})$.

\begin{figure}[ht]
   \epsfxsize=4.5in 
\centerline{\epsffile{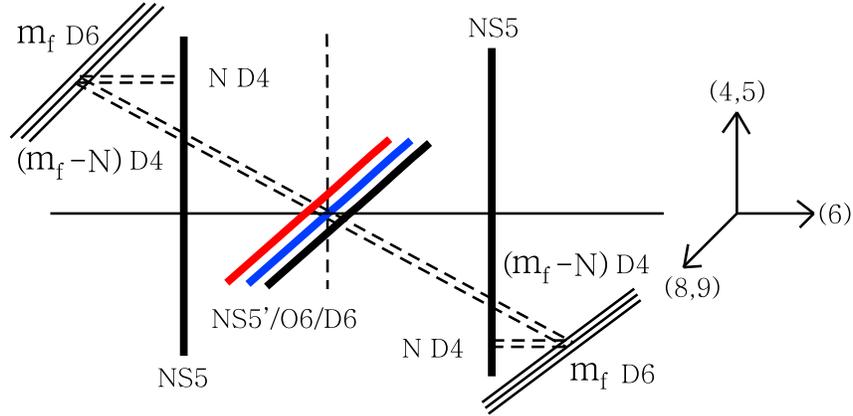}}
   \caption[FIG. \arabic{figure}.]{ 
The nonsupersymmetric minimal energy
brane configuration for the $SU(N=2m_f-N_c+4)$ gauge theory  with 
$a$, $\widetilde{s}$, $q, \widetilde{q}$ as well as 
$\hat{q}$. This can be obtained from the magnetic
description of Figure 2 by putting $\theta=0$ and
$\omega=\frac{\pi}{2}$ 
and with the condition that the number of
D4-branes connecting between D6-branes and NS5'-brane are reduced to a half.
The $(m_f-N)$ D4-branes connecting between
NS5'-brane and D6-branes can move freely along $w$ direction and this
leads to nonzero vacuum expectation values of the meson $M_0$ (\ref{mesons}).
In section 5,
we move each half of the middle NS5'-brane  to the
$ \pm v$ direction holding everything else fixed, instead of moving D6-branes.}
\label{fig3}
\end{figure}

\section{M-theory description of 
nonsupersymmetric meta-stable brane configuration}

The M5-brane lives in (0123) directions and  wraps on a Riemann surface 
inside (4568910) directions. 
The Taub-NUT space (45610) is parametrized by 
two complex variables $(v,y)$ while flat two-dimensions (89) are by
a complex variable $w$.
One of the complex structures of Taub-NUT
space (45610) can be described by embedding it in a three complex dimensional
space
with coordinates $(x,t,v)$. 
The M5-brane curve corresponding to the type IIA brane configuration 
shown in Figure 1, in a background space \cite{Witten} of
$
x t = (-1)^{m_f} v^4 \prod_{k=1}^{m_f} (v^2 -e_k^2)
$
where $e_k$ is the position of the D6-branes in the $v$ direction,
can be characterized by a cubic equation in $t$ \cite{LLL,Ahn07}.
For fixed $x$, the coordinate $t$ corresponds to $y$ and for fixed
$y$, the coordinate $x$
corresponds to $1/t$. 

At nonzero string coupling 
$g_s \neq 0$, the two kinds of NS5-branes can bend due to their interactions
with the D4-branes and D6-branes.
Let us consider the asymptotic behavior such that the rotated curve 
should have at $w \rightarrow \infty$ and $w=0$.
One can read off the behaviors of the left, middle, right NS5-branes
respectively as in \cite{Ahn07}.
Then the behavior of the supersymmetric M5-brane curves can be
written as follows by realizing that 
the role of NS5-brane and NS5'-brane(the behaviors of the coordinates
$v$ and $w$) 
are exchanged into each other:

1. $w \rightarrow \infty$ limit implies
\bea
v \rightarrow \pm m, \quad y \sim    w^{m_f+2} + \cdots \quad
\mbox{NS' asymptotic region}.   
\nonu
\eea

2.  $v \rightarrow \infty$ limit implies
\bea
w & \rightarrow &   0, \quad 
y \sim  v^{2m_f-N_c+4}
 +\cdots
\quad \mbox{$NS_{L}$ asymptotic region}, 
\nonu
\\
w & \rightarrow &  0, \quad  
y \sim v^{N_c}
+\cdots
\quad \mbox{$NS_{R}$ asymptotic region}. 
\nonu
\eea
Along the line of \cite{BGHSS,Ahn06-1,Ahn07}, each half of 
the central NS5'-brane is moved to the
$\pm v$ direction holding everything else fixed, 
instead of moving D6-branes(and their mirrors).
In this process, the corresponding mirrors are placed in appropriate way. 
Since the nonsupersymmetric brane configuration in Figure 3 implies that 
only the NS5'-brane is deformed by turning on the mass for the
quarks, only NS5'-brane is nonholomorphic.
The remaining NS5-brane and D6-branes remain unchanged.
The ansatz for this nonholomorphic 
curve corresponding to the NS5'-brane with
boundary condition above
can be made similarly.

The harmonic function appearing in the Taub-NUT space, sourced by the the left
coincident $m_f$ D6-branes located at $x^6=0$, $O6^{+}$-plane(which is equivalent to 4
D6-branes) 
located at
$x^6=l$, and the right coincident $m_f$ D6-branes located at
$x^6=2l$,
can be written as explicitly, by putting the right charges and the
locations of them,
\bea
V(s) = 1  + 
\frac{m_f R}{\sqrt{f(s)^2 + s^2}}+ 
\frac{4 R}{\sqrt{f(s)^2 + (s-l)^2}}+
\frac{m_f R}{\sqrt{f(s)^2 + (s-2l)^2}}.
\label{harmonic}
\eea
Note that when we compare with the usual ${\cal N}=1$ SQCD with
massive flavors developed in \cite{BGHSS}, the last two terms in (\ref{harmonic})  
coming from the effect of O6-plane and mirrors of
D6-branes in our gauge theory are present newly.   
When half of the middle NS5'-brane is located at $v =\frac{\Delta x}{\ell_s^2}$
from Figure 3(and its mirror appears at opposite value of $v$) and
$
f(s) =  \Delta x
$
\footnote{Remember that the ansatz for the
nonholomorphic curve implies that $x^4=f(s)$ and $x^5=0$ 
and for the mirror of NS5'-brane we should have $f(s) =-\Delta x$
due to the ${\bf Z}_2$ symmetry of O6-plane.}, 
then there exists the equation (A.3) of \cite{BGHSS} which is valid for arbitrary
form for the harmonic function.
Then, this first order differential equation leads to the solution for $g(s)$, 
\bea
g(s)       
\sim \left(s-l + \sqrt{(\Delta
    x)^2 
+ (s-l)^2} \right)^{\frac{2}{N_c}}
 \prod_{j=0}^{1} \left[ s-2lj + \sqrt{(\Delta
    x)^2 
+ (s-2lj)^2} \right]^{\frac{m_f}{2N_c}}
\nonu
\eea
where the $s$-independent integration constant is fixed by 
the boundary condition above classification 2. 
The function $g(s)$ does not vanish even if $\Delta x$ vanishes.
Then $w$ that is proportional to $g(s)$ 
does not vanish and $y$ is not equal to zero. This is not
consistent with the statement that $y=0$ only if $v=0$. 
Therefore, the extra piece in the potential (\ref{harmonic}) doesn't remove
the instability from a new M5-brane mode
occurring  at some point during the continuation to
M-theory description of SQCD.

\section{Summary and outlook}

Let us make some comments on the type IIA brane configuration 
corresponding to the meta-stable vacua for different kinds
of gauge theories. 

$\bullet$ Orthogonal group with a symmetric tensor flavor and massive vectors

As observed in \cite{Ahn07}, it is hard to imagine how the 
meta-stable vacua supersymmetry breaking brane configuration arises because 
there is no extra NS5-brane we need to have. 
The type IIA 
supersymmetry breaking brane configuration of meta-stable vacua 
corresponding to $SU(2N_f-N_c)$ gauge theory with matters 
is shown in
Figure 3 of \cite{Ahn07}. If we go on baryonic branch(i.e., we move a
middle NS5-brane to $x^7$ direction), the
baryon gets an expectation value  along this flat direction
and the gauge group is broken to
$SO(n)$ with a symmetric tensor as well as vectors. 
The superpotential has a quadratic
expression on this symmetric tensor because the D4-branes will cross
the $O6^{+}$-plane in the region giving a projection to the
orthogonal gauge group \cite{BHKL}. Therefore, the type IIA supersymmetry breaking
brane configuration is the Figure 3 of \cite{Ahn07} with a shift of
NS5-brane to $x^7$ direction. The gauge theory analysis on this fact
was given in \cite{ILS} or the flow from mezzanine to orchestra in the
IR of $A_k$ model \cite{BS} corresponds to this example in the gauge theory side. 
In order to obtain higher order term for the
superpotential in terms of the symmetric tensor, we need to insert a
multiple of outer NS5'-branes. 

$\bullet$ Symplectic group with an antisymmetric tensor flavor and
massive fundamental flavors 

From the Figure 3 of \cite{Ahn07} where $O6^{+}$-plane is replaced by 
$O6^{-}$-plane, 
if we move a
middle NS5-brane to $x^7$ direction, the
baryon gets an expectation value along this flat direction. 
Then the gauge group is broken to
$Sp(n)$ with an antisymmetric tensor flavor as well as fundamental
flavors and the superpotential has a quadratic
expression on this antisymmetric tensor because  the D4-branes will cross
the $O6^{-}$-plane in the region which gives a projection to the
symplectic gauge group.
Similarly,  the type IIA supersymmetry breaking
brane configuration is the Figure 3 of \cite{Ahn07} where $O6^{+}$-plane is replaced by 
$O6^{-}$-plane with a shift of
NS5-brane to $x^7$ direction.

$\bullet$ Orthogonal group with an adjoint(antisymmetric tensor) and
massive vectors

From Figure 3 above in this paper, we can move NS5'-brane to $+x^7$ direction. 
The expectation value on this operator in magnetic
dual breaks the dual gauge group $SU(N)$ to $SO(n)$ with an adjoint
as well as vectors and a quadratic superpotential for this adjoint
because the D4-branes will cross
the $O6^{+}$-plane in the region which gives a projection to the
orthogonal gauge group \cite{LLL1,BHKL}. 
Therefore, the type IIA supersymmetry breaking
brane configuration is the Figure 3 in this paper with a shift of
NS5-brane to $+x^7$ direction. The gauge theory analysis on this fact
was given in \cite{ILS} or
 the flow from balcony to mezzanine in the
IR of $A_k$ model \cite{BS} corresponds to this example, in the gauge theory side. 

On the other hand, according to \cite{EGKRS}, this theory can be realized in terms of
$O4^{+}$-plane, NS5-brane, NS5'-brane, D4-branes and D6-branes.
In principle, one can follow the method of \cite{AGM,AGM1} by adding
the corresponding mesonic
deformations and one
expects that one can obtain 
the meta-stable vacua to this theory in the gauge theory side. 

$\bullet$ Symplectic  group with an adjoint(symmetric tensor) and
massive fundamental flavors 

From Figure 3, we can move NS5'-brane to $-x^7$ direction. 
The expectation value on this operator in magnetic
dual breaks the dual gauge group $SU(N)$ to $Sp(n)$ with an adjoint
and a quadratic superpotential for this adjoint
because the D4-branes will cross
the $O6^{-}$-plane in the region which gives a projection to the
symplectic gauge group \cite{LLL1,BHKL}. 
Therefore, the type IIA supersymmetry breaking
brane configuration is the Figure 3 in this paper with a shift of
NS5-brane to $-x^7$ direction. 

Similarly,
one realizes that this gauge theory corresponds to the brane configuration
which
consists of 
$O4^{-}$-plane, NS5-brane, NS5'-brane, D4-branes and D6-branes \cite{EGKRS}.
It is evident that 
one also obtains the corresponding brane configuration from Figure 4 of \cite{Ahn06}     
by inserting all the mirrors with respect to $O4^{-}$-plane appropriately.

Then it is natural to ask what happens for
the product gauge groups. For example, in the gauge theory side
\cite{ILS}, the theory of $SU(N_1) \times SO(N_2)$ with $X$, which
transform as fundamental in $SU(N_1)$ and vector in $SO(N_2)$, a
conjugate symmetric tensor $\widetilde{S}$ in $SU(N_1)$, as well as
fundamentals for $SU(N_1)$ and vectors for $SO(N_2)$ 
can confine either $SU(N_1)$ factor or $SO(N_2)$ factor. For the
latter, the IR duality is given by  
an unitary gauge theory 
with a symmetric tensor flavor and fundamental flavors.
See
also relevant paper \cite{LO} where the matter contents are
different. 

According to the result of \cite{dBO}, by rotating ${\cal N}=2$ brane
configuration, a single smooth supersymmetric 
M5-brane description gives rise to the
monopole and meson expectation values which are matching with the field
theory computation through confining phase superpotential
\cite{EFGIR}. See also the relevant discussions for symplectic and orthogonal gauge
groups
in \cite{Ahn97,Ahn97-1}.
The ${\cal N}=1$ supersymmetric magnetic configuration for this case where 
$k'$ NS5'-branes are connected by $(2N_f-N_c)$ D4-branes to
a single NS5-brane on their right and by a single D4-brane to each of 
$N_f$ D6-branes on their left was given in \cite{EGKRS}.
For $k'=2$ case, the gauge theory analysis was given in \cite{ASY}
where the magnetic dual superpotential corresponding to the electric
superpotential $W=X^3$ with an adjoint $X$
is given by 
$W_{dual} = Y^3 + M_1 \widetilde{q} Y q
+ M_2 \widetilde{q} q$ with a dual adjoint $Y$, dual quarks $q,
\widetilde{q}$ and gauge singlets $M_1 \equiv Q \widetilde{Q}$ 
and $M_2 \equiv Q X \widetilde{Q}$ for electric quarks $Q, \widetilde{Q}$
and an adjoint $X$.
It would be interesting to study whether there exist nonsupersymmetric 
meta-stable vacua.

\vspace{.7cm}

\centerline{\bf Acknowledgments}

I would like to thank A. Hanany and K. Landsteiner for discussions.
This work was supported by grant No.
R01-2006-000-10965-0 from the Basic Research Program of the Korea
Science \& Engineering Foundation.  
I thank KIAS(Korea Institute for 
Advanced Study) for hospitality  where
this work was undertaken.


\begin{thebibliography}{99}

\bibitem{EGK}
  S.~Elitzur, A.~Giveon and D.~Kutasov,
  Phys.\ Lett.\ B {\bf 400}, 269 (1997)

\bibitem{GK}
  A.~Giveon and D.~Kutasov,
  Rev.\ Mod.\ Phys.\  {\bf 71}, 983 (1999)

\bibitem{ISS}
  K.~Intriligator, N.~Seiberg and D.~Shih,
  JHEP {\bf 0604}, 021 (2006)

\bibitem{OO}
  H.~Ooguri and Y.~Ookouchi,
  Phys.\ Lett.\ B {\bf 641}, 323 (2006)

\bibitem{FGU}
  S.~Franco, I.~Garcia-Etxebarria and A.~M.~Uranga,
  JHEP {\bf 0701}, 085 (2007)

\bibitem{BGHSS}
  I.~Bena, E.~Gorbatov, S.~Hellerman, N.~Seiberg and D.~Shih,
  JHEP {\bf 0611}, 088 (2006)

\bibitem{Ahn06-1}
  C.~Ahn,
  Phys.\ Lett.\  B {\bf 647}, 493 (2007)

\bibitem{AGM}
  A.~Amariti, L.~Girardello and A.~Mariotti,
  JHEP {\bf 0612}, 058 (2006)

\bibitem{AGM1}
  A.~Amariti, L.~Girardello and A.~Mariotti,
  [arXiv:hep-th/0701121].

\bibitem{Ahn06}
  C.~Ahn,
  Class.\ Quant.\ Grav.\  {\bf 24}, 1359 (2007)

\bibitem{LLL}
  K.~Landsteiner, E.~Lopez and D.~A.~Lowe,
  JHEP {\bf 9807}, 011 (1998)

\bibitem{Ahn07}
  C.~Ahn,
  [arXiv:hep-th/0701145].

\bibitem{LLL1}
  K.~Landsteiner, E.~Lopez and D.~A.~Lowe,
  JHEP {\bf 9802}, 007 (1998)

\bibitem{BHKL}
  I.~Brunner, A.~Hanany, A.~Karch and D.~Lust,
  Nucl.\ Phys.\ B {\bf 528}, 197 (1998)

\bibitem{EGKT}
  S.~Elitzur, A.~Giveon, D.~Kutasov and D.~Tsabar,
  Nucl.\ Phys.\ B {\bf 524}, 251 (1998)

\bibitem{LL}
  K.~Landsteiner and E.~Lopez,
  Nucl.\ Phys.\ B {\bf 516}, 273 (1998)

\bibitem{ILS}
  K.~A.~Intriligator, R.~G.~Leigh and M.~J.~Strassler,
  Nucl.\ Phys.\ B {\bf 456}, 567 (1995)

\bibitem{ABFK}
  R.~Argurio, M.~Bertolini, S.~Franco and S.~Kachru,
  JHEP {\bf 0701}, 083 (2007)

\bibitem{KOO}
  R.~Kitano, H.~Ooguri and Y.~Ookouchi,
  Phys.\ Rev.\  D {\bf 75}, 045022 (2007)

\bibitem{BH}
  J.~H.~Brodie and A.~Hanany,
  Nucl.\ Phys.\ B {\bf 506}, 157 (1997)

\bibitem{GP}
  A.~Giveon and O.~Pelc,
  Nucl.\ Phys.\ B {\bf 512}, 103 (1998)

\bibitem{ENR}
  J.~Erlich, A.~Naqvi and L.~Randall,
  Phys.\ Rev.\ D {\bf 58}, 046002 (1998)

\bibitem{EJS}
  N.~J.~Evans, C.~V.~Johnson and A.~D.~Shapere,
  Nucl.\ Phys.\ B {\bf 505}, 251 (1997)

\bibitem{HZ}
  A.~Hanany and A.~Zaffaroni,
  Nucl.\ Phys.\ B {\bf 509}, 145 (1998)

\bibitem{Barbon}
  J.~L.~F.~Barbon,
  Phys.\ Lett.\ B {\bf 402}, 59 (1997)

\bibitem{BS}
  J.~H.~Brodie and M.~J.~Strassler,
  Nucl.\ Phys.\ B {\bf 524}, 224 (1998)

\bibitem{HW}
  A.~Hanany and E.~Witten,
  Nucl.\ Phys.\ B {\bf 492}, 152 (1997)

\bibitem{Shih}
  D.~Shih,
  [arXiv:hep-th/0703196].

\bibitem{Witten}
  E.~Witten,
  Nucl.\ Phys.\ B {\bf 500}, 3 (1997)

\bibitem{EGKRS}
  S.~Elitzur, A.~Giveon, D.~Kutasov, E.~Rabinovici and A.~Schwimmer,
  Nucl.\ Phys.\ B {\bf 505}, 202 (1997)

\bibitem{LO}
  E.~Lopez and B.~Ormsby,
  JHEP {\bf 9811}, 020 (1998)

\bibitem{dBO}
  J.~de Boer and Y.~Oz,
  Nucl.\ Phys.\ B {\bf 511}, 155 (1998)

\bibitem{EFGIR}
  S.~Elitzur, A.~Forge, A.~Giveon, K.~A.~Intriligator and E.~Rabinovici,
  Phys.\ Lett.\ B {\bf 379}, 121 (1996)

\bibitem{Ahn97}
  C.~Ahn,
  Phys.\ Lett.\ B {\bf 426}, 306 (1998)

\bibitem{Ahn97-1}
  C.~Ahn, K.~Oh and R.~Tatar,
  J.\ Geom.\ Phys.\  {\bf 28}, 163 (1998)

\bibitem{ASY}
  O.~Aharony, J.~Sonnenschein and S.~Yankielowicz,
  Nucl.\ Phys.\ B {\bf 449}, 509 (1995)











\end{thebibliography}
\end{document}